# The Inability of the White-Juday Warp Field Interferometer to Spectrally Resolve Spacetime Distortions

Jeffrey S. Lee[1]
Gerald B. Cleaver[2]

[1,2]Early Universe Cosmology and Strings Group, a Division of the Center for Astrophysics, Space Physics, and Engineering Research
[2]Department of Physics
Baylor University
One Bear Place
Waco, TX 76706

[1,2]Icarus Interstellar
Anchorage, Alaska

[1]Crescent School
Toronto, Ontario

Jeff_Lee@baylor.edu
Gerald_Cleaver@baylor.edu

This paper contends that the spacetime distortions resulting from the experimentally obtainable electric field of a parallel plate capacitor configuration cannot be detected by the White-Juday Warp Field Interferometer [1]. Any post-processing results indicating a vanishing, non-zero difference between the charged and uncharged states of the capacitor are due to local effects rather than spacetime perturbations.

## Introduction

Located in the Eagleworks Laboratory at the Johnson Space Center in Houston, Texas, the White-Juday Warp Field Interferometer (WJWFI), which is a modified, seismically-isolated Fabry-Pérot interferometer, has been developed to detect spacetime distortions created by a $\sim 10^6$ V·m$^{-1}$ static electric field. The interferometer employs a 6328 Å HeNe laser, in which one of the two beams passes between two electrically charged parallel plates. The beams are recombined on a CCD array.

However, the spacetime distortions produced by such an electric field are exceptionally below the detection threshold of all present-day interferometry techniques.

Additionally, an analysis of refractive index variations, due to plausible air temperature differences in the laboratory, was conducted, and the resulting beam refraction is shown to be potentially above the lower limit of detectability of the WJWFI.

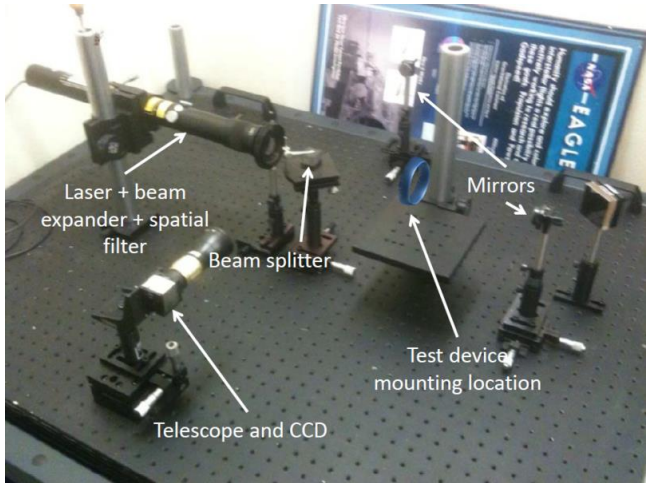

Figure 1: The White-Juday Warp Field Interferometer[i].

## The Spacetime Distortions due to the Electric Field

The exact electric field strength between the plates being used in this experiment has not been specified in the literature. However, in this paper, we assume a generous $10^6$ V·m$^{-1}$ static electric field created by a 100 kV potential difference across a 0.1 m air gap. Therefore, the energy density $(u_E)$ produced is:

$$u_E = \frac{\varepsilon_o V^2}{2d^2} = 4.4 \text{ J} \cdot \text{m}^{-3} \quad (1)$$

Evaluation of the full normalized stress energy tensor is not necessary to affirm that the time-independent spacetime distortion, which is determined from (2) and is $9.2 \times 10^{-43}$ m$^{-2}$, cannot be resolved by any known experimental technique.

$$G_{\mu\nu} = \frac{8\pi G}{c^4} T_{\mu\nu} \quad (2)$$

Additionally, an Alcubierre warp bubble can be formed only by negative energy density and/or negative pressure [2], and it is improbable that this experimental technique could be seamlessly extrapolated for the case of -$\rho$ terms in $T_{\mu\nu}$.

For the White-Juday Warp Field Interferometer to attempt to detect the microlensing caused by the above 4.4 J·m$^{-3}$ electric field with a 0.1 m radius and a $L$ = 10 m path length, the following results. From (3), (4), and (5) respectively, the Schwarzschild radius $(R_S)$ is $1.92 \times 10^{-23}$ m, the Einstein radius $(R_E)$ is $8.76 \times 10^{-12}$ m, and the angular diameter $(\theta)$ is $1.81 \times 10^{-6}$ arcseconds, which is 3 orders of magnitude below the maximum discernable resolution of approximately $10^{-3}$ arcseconds [3].

---

[i] From: http://ntrs.nasa.gov/archive/nasa/casi.ntrs.nasa.gov/20130011213.pdf

$$R_S = \frac{2GM}{c^2} \tag{3}$$

$$R_E = \sqrt{\frac{2L}{R_S}} \tag{4}$$

$$\theta = \frac{2R_E}{L} \tag{5}$$

The WJWFI is totally incapable of detecting the minute distortions of spacetime produced by a 4.4 J·m$^{-3}$ electric field. The static electric field of equivalent radius required to achieve the microlensing detection threshold would be ~$10^{12}$ V·m$^{-1}$. Therefore, any vanishing non-zero difference between the charged and uncharged states of the plates is clearly due other factors.

## Air Refraction

The dependence of the index of refraction of air on the vacuum wavelength of incident EM radiation is given by:

$$(n-1) \times 10^8 = 8340.78 + \left[\frac{2.405640 \times 10^6}{130 - \lambda^{-2}}\right] + \left[\frac{1.5994 \times 10^4}{38.9 - \lambda^{-2}}\right] \tag{6}$$

(7) accounts for arbitrary pressures and the presence of water vapor, and is plotted in Figure 2.

$$(n-1) \times 10^8 = \left\{8340.78 + \left[\frac{2.405640 \times 10^6}{130 - \lambda^{-2}}\right] + \left[\frac{1.5994 \times 10^4}{38.9 - \lambda^{-2}}\right]\right\} \times \left(\frac{p}{720.775}\right)\left[\frac{1 + p(0.817 - 0.0133T) \times 10^{-6}}{1 + 0.0036610T}\right] - f\left[5.722 - \frac{0.0457}{\lambda^2}\right] \tag{7}$$

where $n$ is the index of refraction, $\lambda$ is the vacuum wavelength (m), $T$ is the temperature (Celsius), $p$ is the total pressure (torricelli), and $f$ is the partial pressure of water vapor (torricelli) [4].

For this analysis, the atmosphere in the Eagleworks laboratory is assumed to be dry, $CO_2$-free air with the molar composition, pressure and approximate temperature as shown in Table 1.

| Composition | Molar Percentage |
|---|---|
| $N_2$ | 78.09 |
| $O_2$ | 20.95 |
| Ar | 0.93 |
| $CO_2$ | 0.03 |

Table 1: Composition and molar percentage of the components of dry air ($T$ = 288.16 K, $P$ = 1013.25 mb) **[4]**.

By dividing the Edlén equation [5] by 1.000162, the effect of carbon dioxide on refraction is removed [4].

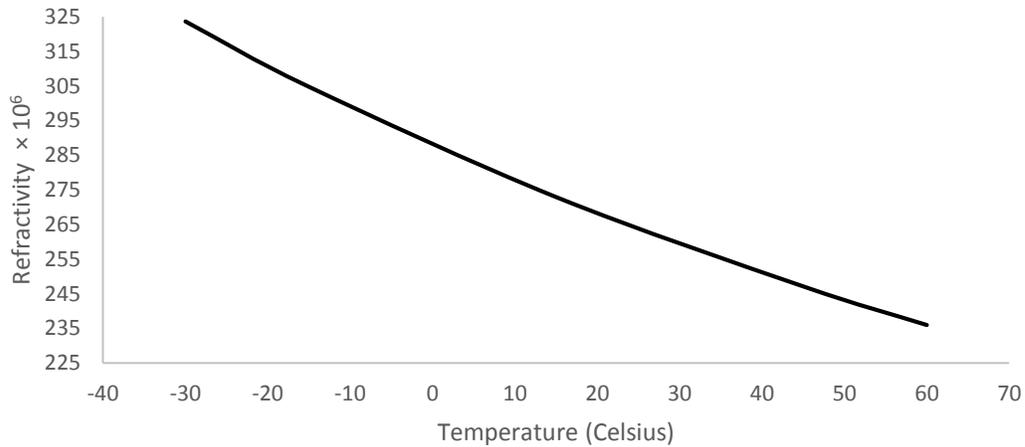

Figure 2: Refractivity versus temperature ($p$ = 1013.25 mb), $\lambda$ = 6328 Å, RH%[ii] = 0).

For a first order model, a 6328 Å laser beam is incident at 0.1° upon the interface between parcels of still air at 20° C and 21° C; the resulting difference in the refraction angles at the interface is $10^{-7}$ °. When permitted to travel along a 2.0 m post-interface path, the lateral beam deviation, as a result of refraction, is 31.7 Å, and is thus significantly below the threshold of detectability of the WJWFI. Only for angles of incidence greater than approximately 17° would the lateral deviation of the beam be comparable to, or greater than, the laser wavelength.

Unless the WJWFI experiment is performed in vacuua, refraction-induced beam divergences along the arms of the interferometer cannot be eliminated and could be sporadically evident.

## Conclusion

The White-Juday Warp Field Interferometer has been demonstrated to be incapable of resolving the minute distortions of spacetime created by a $10^6$ V·m$^{-1}$ electric field.

Variations in temperature were shown to produce potentially detectable changes in the refractive index of air, which could result in occasional spurious interference fringes. Although a more rigorous model, which considers a time- and spatially-changing index of refraction gradient along the interferometer arm,

---

[ii] Relative humidity.

would result in a smaller lateral beam deviation, the purpose for which the WJWFI is intended has been shown to be unachievable.

Thus, were any signals to appear in the White-Juday Warp Field Interferometer, they would most often be attributable to either electronic noise or the classical electrodynamics interaction between the ionized air between the plates and the electromagnetic radiation of the laser.

## Works Cited

[1] H. White, *JBIS,* vol. 66, pp. 242-247, 2013.

[2] M. Alcubierre, "The warp drive: hyper-fast travel within general relativity.," *Journal of Classical and Quantum Gravity,* vol. 11, pp. L73-L77, 1994; arXiv:0009013 [gr-qc].

[3] M. B. Bogdanov, "Ultra-high Angular Resolution by Gravitational Microlensing," September 2000; arxiv.org:0009/0009228 [astro-ph].

[4] J. C. Owens, "Optical Refractive Index of Air: Dependence on Pressure, Temperature and Composition," *Applied Optics,* vol. 6, no. 1, pp. 51-59, January 1967.

[5] B. Edlén, *Metrologia,* vol. 2, no. 71, 1966.